\documentclass[preprint,superscriptaddress,nofootinbib,showkeys,showpacs,tightenlines]{revtex4}
\usepackage{graphicx}
\usepackage{dcolumn}
\usepackage{bm}
\newcommand{\be}{\begin{equation}}
\newcommand{\ee}{\end{equation}}
\newcommand{\bee}{\begin{eqnarray}}
\newcommand{\eee}{\end{eqnarray}}
\newcommand{\eq}{\end{quote}}
\newcommand{\nn}{\nonumber}
\newcommand{\Slash}[1]{\ooalign{\hfil/\hfil\crcr$#1$}}

\begin{document}      

\preprint{PNU-NTG-02/2004}
\preprint{PNU-NURI-01/2004}
\title{Production of the pentaquark $\Theta^+$ in $np$ scattering}
\author{Seung-Il Nam}
\email{sinam@rcnp.osaka-u.ac.jp}
\affiliation{Research Center for Nuclear Physics (RCNP), Ibaraki, Osaka
567-0047, Japan}
\affiliation{Department of
Physics and Nuclear physics \& Radiation technology Institute (NuRI), Pusan National University, Busan 609-735, Korea}

\author{Atsushi Hosaka}
\email{hosaka@rcnp.osaka-u.ac.jp}
\affiliation{Research Center for Nuclear Physics (RCNP), Ibaraki, Osaka
567-0047, Japan}

\author{Hyun-Chul Kim}
\email{hchkim@pusan.ac.kr}
\affiliation{Department of
Physics and Nuclear physics \& Radiation technology Institute (NuRI), Pusan National University, Busan 609-735, Korea}

\date{\today}
\begin{abstract}
We study $np\rightarrow \Lambda\Theta^{+}$ and $np\rightarrow
\Sigma^{0}\Theta^{+}$ processes for both of the positive and negative
parities of the $\Theta^{+}$.  Employing the effective chiral
Lagrangians for the $KNY$ and $K^*NY$ interactions, we calculate
differential cross sections as well as total cross sections for the
$np\rightarrow \Sigma^0 \Theta^+$ and $np\rightarrow \Lambda\Theta^+$
reactions.  The total cross sections for the positive-parity
$\Theta^+$ turn out to be approximately ten times larger than those
for the negative parity $\Theta^+$ in the range of the CM energy
$\sqrt{s}_{\rm th}\le \sqrt{s}\le 3.5\, {\rm GeV}$.  The results are
rather sensitive to the mechanism of $K$ exchanges in the $t$ --
channel.    
\end{abstract}

\pacs{13.75.-n, 13.75.Cs, 12.39.Mk}
\keywords{$\Theta^{+}$ baryon, Parity, Neutron-proton interaction}

\maketitle

\section{introduction}
Since the experimental finding of the lightest {\em pentaquark} baryon
$\Theta^+$~\cite{Nakano:2003qx} motivated by the work of
Ref.~\cite{Diakonov:1997mm}, the physics of the pentaquark states has
been a hot issue.  The DIANA~\cite{Barmin:2003vv}, CLAS~\cite{clas},
SAPHIR~\cite{Barth:2003es}, HERMES~\cite{hermes}, SVD~\cite{svd}
collaborations and the reanalysis of neutrino data~\cite{neutrino} have  
confirmed its existence.  The $\Theta^+$ has unique features: 
It has a relatively small mass and a very narrow width.  The exotic
$\Xi$ states found recently by the NA49 collaboration~\cite{NA49}
share the features similar to the $\Theta^+$.  While a great amount of
theoretical effort has been put into understanding properties of the
$\Theta^+$~\cite{quark,chiral,chiral2,bound,qsr,lqcd,reaction}, there is no
consensus in determining the parity of the $\Theta^+$.  For example,
chiral models predict the parity of the $\Theta^+$ to be
positive~\cite{chiral}, whereas the lattice QCD and the QCD sum rule
prefer the negative parity~\cite{lqcd,qsr}.  

Many works have suggested different ways of determining the parity of the 
$\Theta^+$~\cite{Nakayama:2003by,Zhao:2003bm,Yu:2003eq,Carlson:2003xb,
Thomas:2003ak,Hanhart:2003xp,Nam:2004qy,Mehen:2004dy,Rekalo:2004kb},
among which Thomas {\em et al.}~\cite{Thomas:2003ak} have proposed an
unambiguous method to determine the parity of the $\Theta^+$ via
polarized proton-proton scattering at and just above threshold of the
$\Theta^+$ and $\Sigma^+$: If the parity of the $\Theta^+$ is positive,
the reaction is allowed at the threshold region only when the total spin of
the two protons is $ S = 0$, while negative the reaction is
allowed only when $ S = 1$. Hence it is very challenging to measure such a 
process experimentally~\cite{private1}.  Triggered by
Thomas {\em et al.}~\cite{Thomas:2003ak}, Hanhart {\em et
  al.}~\cite{Hanhart:2003xp} have extended the work of 
Ref.~\cite{Thomas:2003ak} to determine the parity of the $\Theta^+$,
asserting that the sign of the spin correlation function $A_{xx}$
agrees with the parity of the $\Theta^+$ near threshold.  Similarly,   
Rekalo and Tomasi-Gustafsson~\cite{Rekalo:2004kb} have put forward 
methods for the determination of the parity of the $\Theta^+$ by
measuring the spin correlation coefficients in three different
reactions, {\em i.e.} $pn\rightarrow \Lambda\Theta^+$,
$pp\rightarrow \Sigma^+  \Theta^+$, and $pp\rightarrow \pi^+ 
\Lambda \Theta^+$.  Thus, it seems that the $NN$ reactions provide
a promising framework to determine the parity of the $\Theta^+$.  The
present authors have performed the calculation of the cross sections
of the reaction $\vec p \vec p \rightarrow \Sigma^+
\Theta^+$ near the production threshold~\cite{Nam:2004qy}, finding
that the cross sections for the allowed spin configuration are estimated
to be of order of one microbarn for the positive parity $\Theta^+$ and 
about one tenth microbarn for the negative parity $\Theta^+$
in the vicinity of threshold, where the S-wave component dominates.

There exist already investigations on the production of
the $\Theta^+$ in the $NN$
interaction~\cite{Polyakov:1999da,Liu:2003rh,Oh:2003gj,
Bleicher:2004is}.  Refs.~\cite{Liu:2003rh,Oh:2003gj} are 
concerned with the prediction of the total cross sections and
Ref.~\cite{Bleicher:2004is} has explored the $\Theta^+$ production in
high-energy $pp$ scattering.  In the present work, we want to
investigate the $np\rightarrow \Lambda\Theta^{+}$ and $np\rightarrow
\Sigma^{0}\Theta^{+}$ processes with both of the positive and negative
parities considered.  

The present paper is organized as follows: In Section II, we
shall compute the relevant invariant amplitudes from which the total
and differential cross sections can be derived.  In the subsequent
section, we shall present the numerical results and discuss them.  In   
the last section, we shall summarize and draw a conclusion.

\section{Effective Lagrangians and amplitudes}
\begin{figure}[tbh]
\begin{tabular}{c}
\resizebox{15cm}{3cm}{\includegraphics{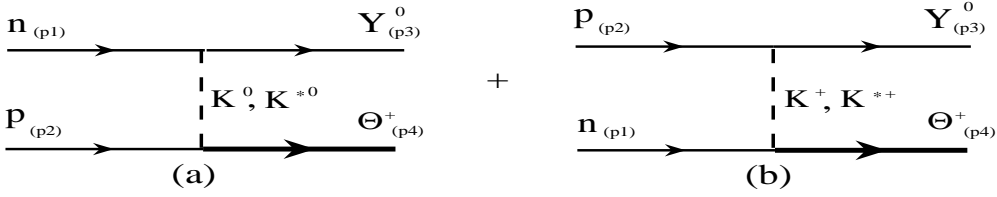}}
\end{tabular}
\caption{Schematic diagrams for the  $np\rightarrow Y^{0}\Theta^{+}$
reaction} 
\label{nmset0}
\end{figure}

The pertinent schematic diagrams for the $np\rightarrow Y^{0}\Theta^{+}$
reaction are drawn in Fig.\ref{nmset0}. 
At the tree level we can consider Born diagrams of pseudoscalar
$K$ and vector $K^{*}$ exchanges.  As mentioned 
before, we treat the reactions in the case of positive- and
negative-parity $\Theta^{+}$.  We distinguish the positive-parity
$\Theta^+$ from the negative-parity one by expressing them as
$\Theta^{+}_{+}$ and $\Theta^{+}_{-}$, respectively.  

We start with the following effective Lagrangians:
\bee
\mathcal{L}_{KNY}&=&-ig_{KNY}\bar{Y}\gamma_{5}K^{\dagger}N,\nn\\
\label{knyl}
\mathcal{L}_{KN\Theta_{\pm}}&=&-ig_{KN\Theta_{\pm}}\bar{\Theta}_{\pm}\Gamma_{5}KN,     
\label{kntl} \cr
\mathcal{L}_{VNY}&=&-g_{VNY}\bar{Y}\gamma_{\mu}
V^{\mu}N-\frac{g^{T}_{VNY}}{M_{Y}+M_{N}}\bar{Y}
\sigma_{\mu\nu}\partial^{\nu}V^{\mu}N,\nn\\ 
\mathcal{L}_{VN\Theta}&=&-g_{VN\Theta_{\pm}}
\bar{\Theta}_{\pm}\gamma_{\mu}\bar{\Gamma}_{5}V^{\mu}
N-\frac{g^{T}_{VN\Theta_{\pm}}}{M_{\Theta}+M_{N}}
\bar{\Theta}_{\pm}\sigma_{\mu\nu}\bar{\Gamma}_{5}\partial^{\nu}V^{\mu}N,
\eee  
where $Y$, $K$, $N$, $\Theta$, and $V$ stand for the hyperon
($\Sigma^{0}$ and $\Lambda$), kaon, nucleon, $\Theta^{+}$, and vector
meson fields, respectively.  In order to take into account different
parities for the $\Theta^{+}$ in the reactions, we introduce
$\Gamma_{5} = \gamma_{5}$ for the $\Theta^{+}_{+}$ and $\Gamma_{5} =
{\bm 1}_{4\times 4} $ for the $\Theta^{+}_{-}$.  $\bar{\Gamma}_{5}$ 
designates $\Gamma_{5}\gamma_{5}$.  The isospin factor is included in
$Y$.  The $KN\Theta$ coupling constant can be determined, if we know
the deacy width $\Gamma_{\Theta\rightarrow KN}$. If we choose
$\Gamma_{\Theta\rightarrow KN} = 15 \,{\rm MeV}$ together with
$M_{\Theta} = 1540\, {\rm MeV}$~\cite{Nakano:2003qx}, we find that 
$g_{KN\Theta_{+}^{+}}$ = 3.78 and $g_{KN\Theta_{-}^{+}}$ = 0.53.  If
one takes a different width for $\Gamma_{\Theta\rightarrow KN}$, the 
coupling constant scales as a square root of the width.  As for
the unknown coupling constant $g_{K^{*}N\Theta}$, we follow
Ref.~\cite{Liu:2003hu}, {\em i.e.}, $g_{K^{*}N\Theta}=\pm
|g_{KN\Theta}|/2$.  The tensor coupling constant
$g^{T}_{K^{*}N\Theta}$ is then fixed as follows:
$g^{T}_{K^{*}N\Theta}=\pm |g_{KN\Theta}|$ as in
Ref.~\cite{Nam:2004qy}.  Since the sign of the coupling constants
cannot be fixed by SU(3) symmetry, we shall use both
signs~\cite{Liu:2003hu}.  We employ the values of the 
$KNY$ and $K^{*}NY$ coupling constants referring to those from the new
Nijmegen potential~\cite{stokes} as well as from the J\"ulich--Bonn
potential~\cite{Reuber:ip} as summarized in Table.~\ref{table1}.
\begin{table}[tbh]
\begin{tabular}{c|cccccc}
&$g_{KN\Lambda}$ &$g_{K^{*}N\Lambda}$ & $g^{T}_{K^{*}N\Lambda}$ &
$g_{KN\Sigma}$ & $g_{K^{*}N\Sigma}$ &$g^{T}_{K^{*}N\Sigma}$\\
\hline
Nijmegen & $-$13.26 & $-$5.19 & $-$13.12 & 3.54 & $-$2.99 & 2.56 \\
J\"ulich--Bonn & $-$18.34 & $-$5.63 & $-$18.34 & 5.38 & $-$3.25 & 7.86
\end{tabular}
\caption{The coupling constants}
\label{table1}
\end{table} 
 
The invariant Feynman amplitudes corresponding to Fig.~\ref{nmset0}
are obtained as follows:
\small
\bee
i\mathcal{M}&=&\Bigg[i\frac{F^{2}(q^{2})g_{KYN}g_{KN\Theta_{\pm}}}{q^{2}-M^{2}_{K}}
\bar{u}(p_{4})\Gamma_{5}u(p_{2})\bar{u}(p_{3})\gamma_{5}u(p_{1})\nn\\
&+&i\frac{F^{2}(q^{2})g_{K^{*}YN}g_{K^{*}N\Theta_{\pm}}}{q^{2}-M^{2}_{K^{*}}}
\big(\bar{u}(p_{4})\gamma^{\mu}\bar{\Gamma}_{5}u(p_{2})\bar{u}(p_{3})
\gamma_{\mu}u(p_{1})-\frac{1}{M^{2}_{K^{*}}}\bar{u}(p_{4})
\Slash{q}\bar{\Gamma}_{5}u(p_{2})\bar{u}(p_{3})\Slash{q}u(p_{1})\big)\nn\\
&-&i\frac{F^{2}(q^{2})g^{T}_{K^{*}YN}g_{K^{*}N\Theta_{\pm}}}{2(M_{N}+M_{Y})
\left(q^{2}-M^{2}_{K^{*}}\right)}\bar{u}(p_{4})\gamma^{\mu}
\bar{\Gamma}_{5}u(p_{2})\bar{u}(p_{3})(\gamma_{\mu}\Slash{q}-
\Slash{q}\gamma_{\mu})u(p_{1})\nn\\
&+&i\frac{F^{2}(q^{2})g_{K^{*}YN}g^{T}_{K^{*}N\Theta_{\pm}}}
{2(M_{N}+M_{\Theta})\left(q^{2}-M^{2}_{K^{*}}\right)}
\bar{u}(p_{4})\bar{\Gamma}_{5}(\gamma^{\mu}\Slash{q}-\Slash{q}
\gamma^{\mu})u(p_{2})\bar{u}(p_{3})\gamma_{\mu}u(p_{1})\nn\\
&-&i\frac{F^{2}(q^{2})g^{T}_{K^{*}YN}g^{T}_{K^{*}N\Theta_{\pm}}}
{4(q^{2}-M^{2}_{K^{*}})(M_{N}+M_{Y})(M_{N}+M_{\Theta})}
\bar{u}(p_{4})(\gamma^{\mu}\Slash{q}-\Slash{q}\gamma^{\mu})
\bar{\Gamma}_{5}u(p_{2})\bar{u}(p_{3})(\gamma_{\mu}\Slash{q}-\Slash{q}
\gamma_{\mu}\Slash{q})u(p_{1})\Bigg]\nn\\
&+&\Big[p_{1}\leftrightarrow p_{2}\Big], 
\label{amplitude}
\eee  
\normalsize   
where $q$ = $p_{1}-p_{3}$.  In order to compute the cross sections for
these reactions, we need the form factors at each vertex to take into
account the extended size of hadrons.  For the Nijmegen potential we
introduce the monopole-type 
form factor~\cite{Machleidt:hj} in the form of 
\bee
F(q^{2}) = \frac{\Lambda^{2}-m^{2}}{\Lambda^{2}-t},
\label{ff1}
\eee
where $m$ and $t$ are the meson mass and a squared four momentum
transfer, respectively. The value of the cutoff parameter is 
taken to be 1.0 GeV for the parameter set of the Nijmegen 
potential~\cite{Nam:2004qy}. As for that of the J\"ulich--Bonn
potential, we make use of the following form factor taken from
Ref.~\cite{Reuber:ip}:  
\bee
F(q^{2}) = \frac{\Lambda^{2}-m^{2}}{\Lambda^{2}+|{\bf q}|^{2}},
\label{ff2}
\eee
where $|{\bf q}|$ is the three momentum transfer.  In this case, we
take different values of the cutoff masses for each $KNY$ vertex as
follows~\cite{Reuber:ip}: $\Lambda_{KN\Theta} = 
\Lambda_{K^{*}N\Theta} = 1.0\, {\rm GeV}$, $\Lambda_{KN\Lambda}=1.2\,
{\rm GeV}$,  $\Lambda_{K^{*}N\Lambda}=2.2\, {\rm GeV}$,
$\Lambda_{KN\Sigma}=2.0\,{\rm GeV}$, and
$\Lambda_{K^{*}N\Sigma}=1.07\, {\rm GeV}$.

\section{Numerical results and discussion}
\begin{figure}[tbh]
\begin{tabular}{c}
\resizebox{8cm}{5cm}{\includegraphics{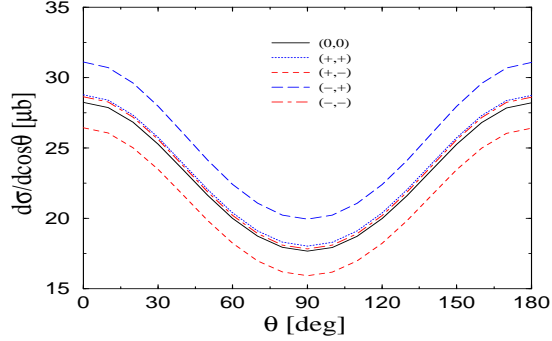}}
\end{tabular}
\caption{The differential cross sections for the reaction 
$np\rightarrow\Lambda\Theta^{+}_{+}$  at $\sqrt{s}$ = 2.7 GeV with
five different combinations of the signs of the $K^{*}N\Theta$
coupling constants as labeled by (sgn($g_{K^*N\Theta}$), 
sgn($g_{K^*N\Theta}^T)$).  The parameter
set of the Nijmegen potential with the cutoff parameter $\Lambda$ =
1.0 GeV is employed.}  
\label{nmset2}
\end{figure}

In this section, we present the total and differential cross sections 
for the reactions $np\rightarrow \Lambda^0 \Theta^+$ and
$np\rightarrow \Sigma^0 \Theta^+$ with two different parities of
$\Theta^{+}$.  We first consider the case of parameter set of the
Nijmegen potential.   
\begin{figure}[tbh]
\begin{tabular}{cc}
\resizebox{8cm}{5cm}{\includegraphics{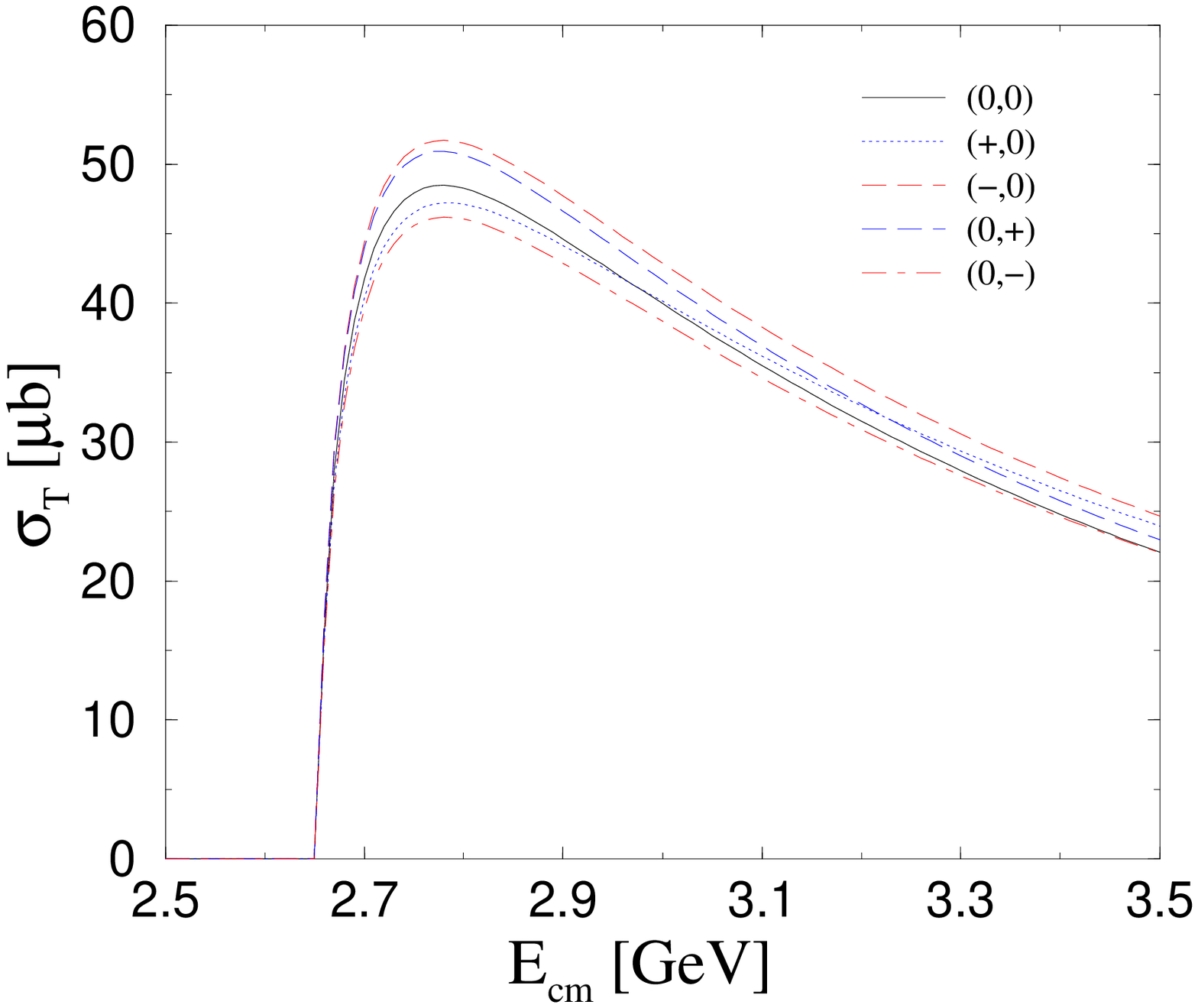}}
\resizebox{8cm}{5cm}{\includegraphics{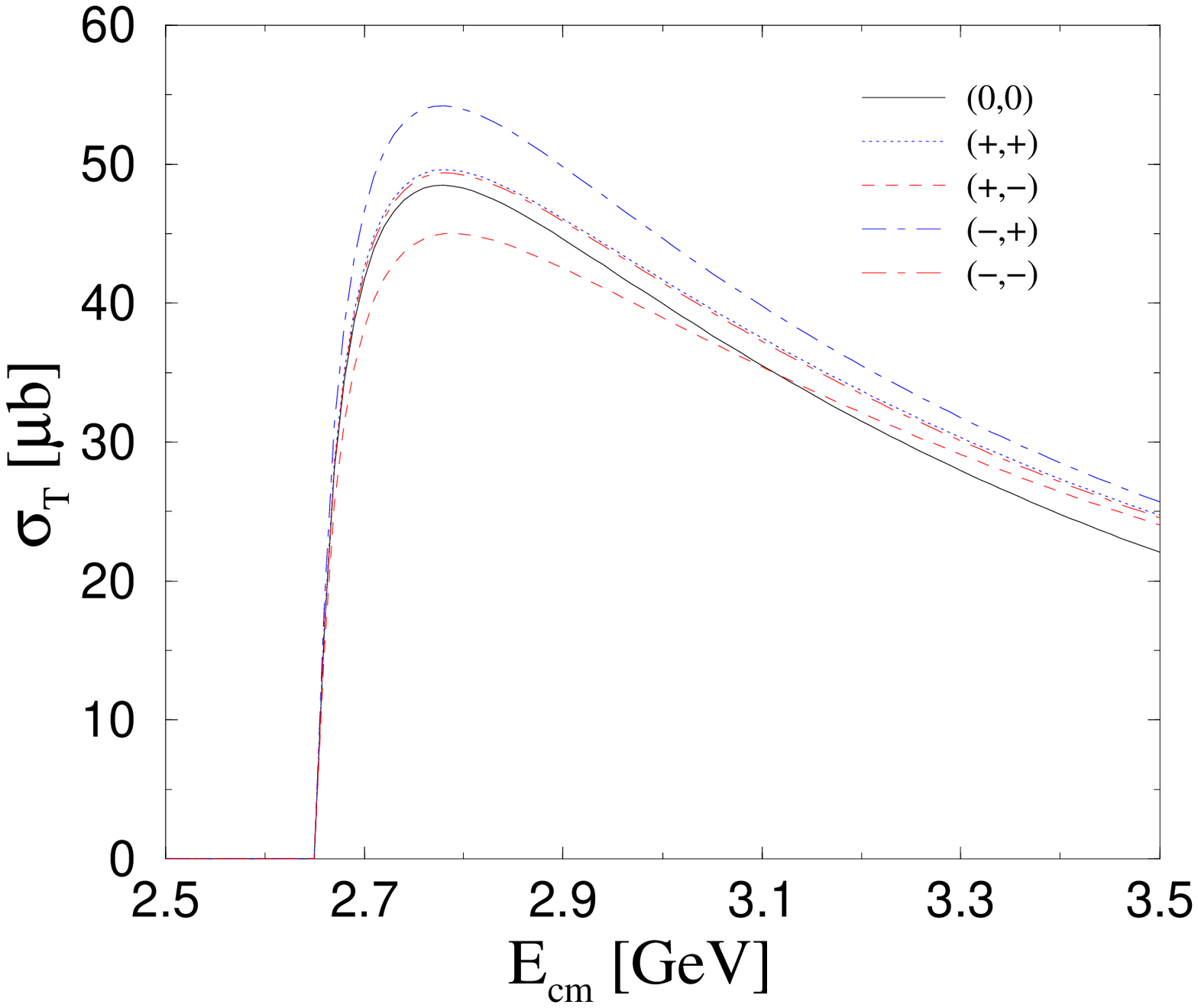}}
\end{tabular}
\caption{The total cross sections of
$np\rightarrow\Lambda\Theta^{+}_{+}$ with ten different
combinations of the signs of the $K^{*}N\Theta$ coupling constants
which are labeled 
by (sgn($g_{K^*N\Theta}$), sgn($g_{K^*N\Theta}^T)$).  The parameter
set of the Nijmegen potential with the cutoff parameter $\Lambda$ =
1.0 GeV is employed.} 
\label{nmset1}
\end{figure}   
In Fig.~\ref{nmset1}, we draw the total cross sections of
$np\rightarrow\Lambda\Theta^{+}_{+}$ for different signs of the
coupling constants, which are labeled as (sgn($g_{K^*N\Theta}$),
sgn($g_{K^*N\Theta}^T)$).  We compare the results from ten different
combinations of the signs.  As shown in Fig.~\ref{nmset1}, the
dependence on the signs is rather weak.  Moreover, we find
that the contribution from $K^*$ exchange is very tiny.  The average
total cross section is obtained as 
$\sigma_{np\rightarrow\Lambda\Theta^{+}_{+}}\sim$  40 $\mu b$ 
in the range of the CM energy $\sqrt{s}_{\rm th} \le \sqrt{s} \le 
3.5$ GeV, where $\sqrt{s}_{\rm th}$ = 2656 MeV. 
Since the angular distribution for all reactions is with a similar 
shape, we show the results only for the case of
$np\rightarrow\Lambda\Theta^{+}_{+}$ in Fig.~\ref{nmset2}. 

\begin{figure}[tbh]
\resizebox{8cm}{5cm}{\includegraphics{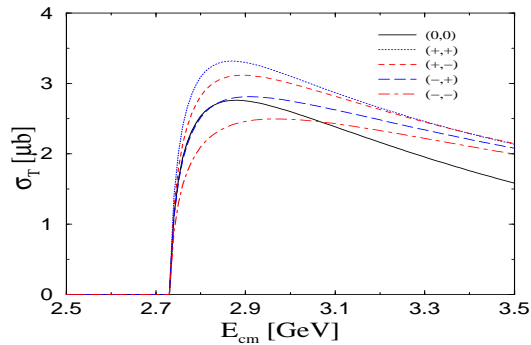}}
\caption{The total cross sections for the reaction
$np\rightarrow\Sigma^{0}\Theta^{+}_{+}$.  The parameter
set of the Nijmegen potential with the cutoff parameter $\Lambda$ = 
1.0 GeV is employed.  The notations are the same as in
Fig.~\ref{nmset2}.}    
\label{nmset22}
\end{figure}    

In Fig.~\ref{nmset22}, we draw the total cross sections for the
reaction $np\rightarrow\Sigma^{0}\Theta^{+}_{+}$.  We find that they
are about ten times smaller than those for the reaction 
$np\rightarrow\Lambda\Theta^{+}_{+}$.  The corresponding average
total cross section is found to be     
$\sigma_{np\rightarrow\Sigma^0\Theta^{+}_{+}}\sim$  2.0 $\mu b$ in the
range of the CM energy $\sqrt{s}_{\rm th} \le \sqrt{s} \le  
3.5$ GeV, where $\sqrt{s}_{\rm th}$ = 2733 MeV. It can be easily
understood from the fact that the ratio of the coupling constants
$|g_{KN\Lambda}/g_{KN\Sigma}| = 3.74$ is rather large and   
the contribution from $K$-exchange is dominant.  
\begin{figure}[tbh]
\begin{tabular}{cc}
\resizebox{8cm}{5cm}{\includegraphics{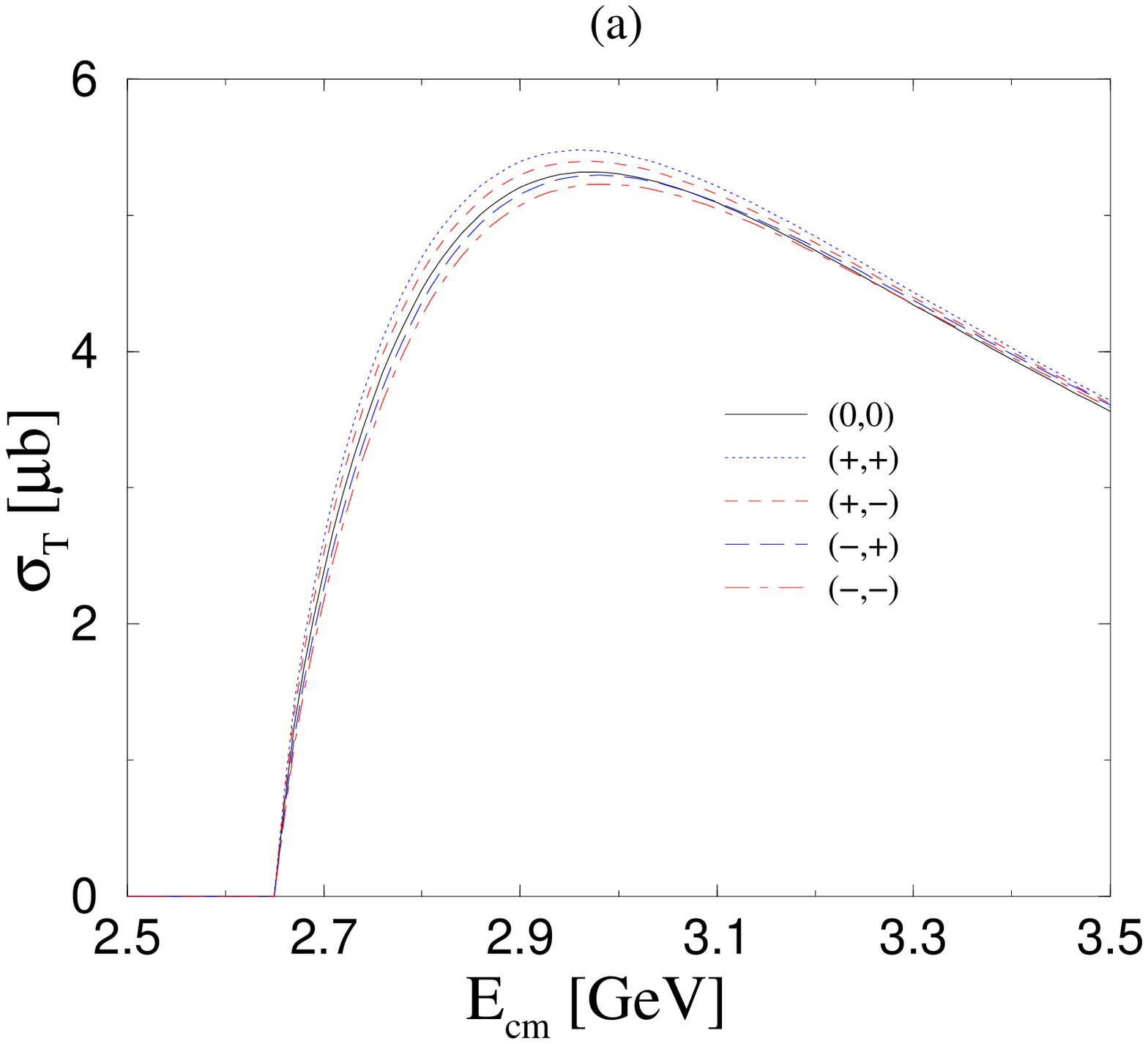}}
\resizebox{8cm}{5cm}{\includegraphics{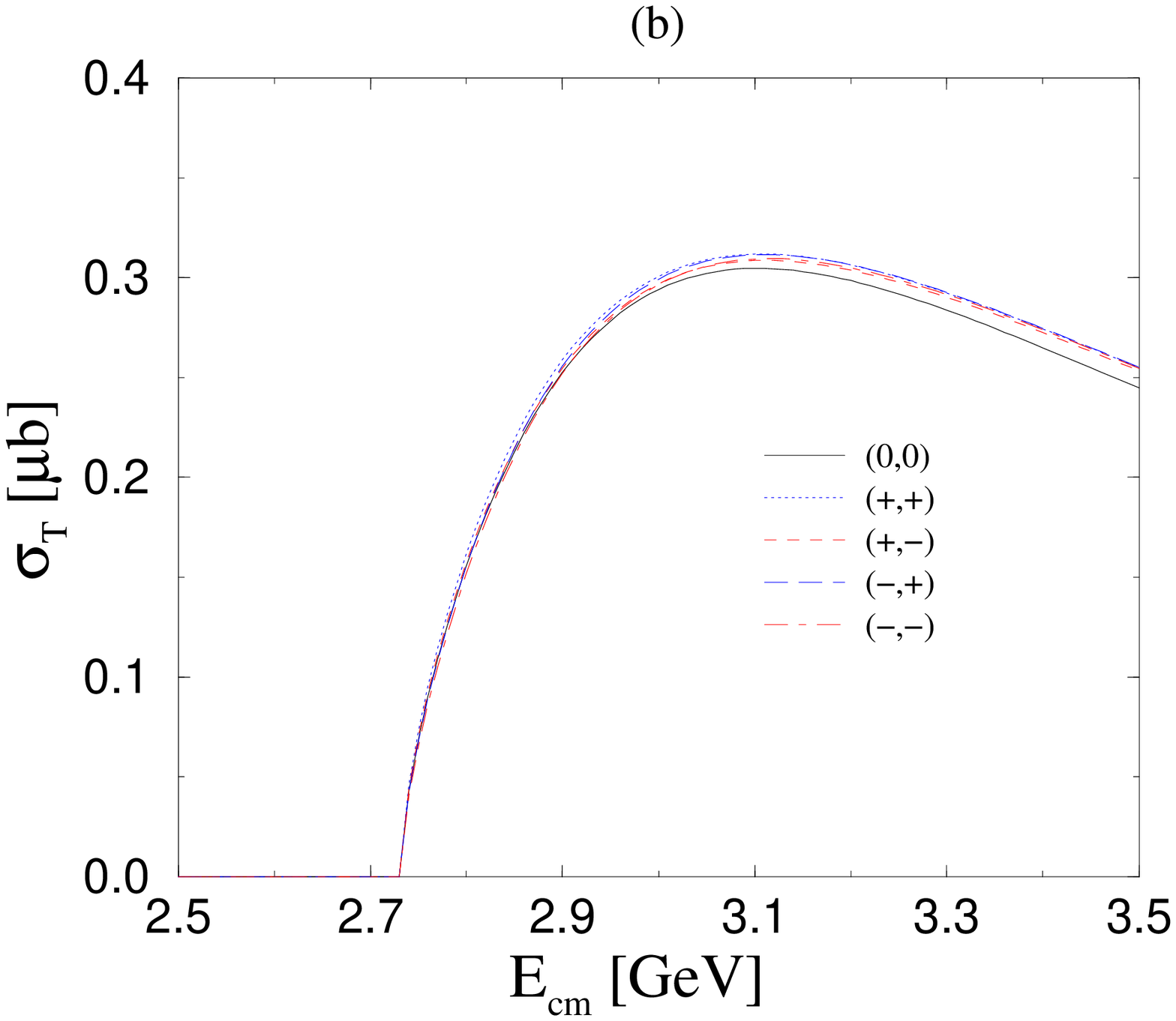}}
\end{tabular}
\caption{The total cross sections of
$np\rightarrow\Lambda\Theta^{+}_{-}$ in the left panel (a) and
$np\rightarrow\Sigma^{0}\Theta^{+}_{-}$ in the right panel (b).  The
parameter set of the Nijmegen potential with the cutoff parameter
$\Lambda$ = 1.0 GeV is employed.  The notations are the same as in
Fig.~\ref{nmset2}.}   
\label{nmset3}
\end{figure}   

As for the negative parity $\Theta^{+}$, we show the results in
Fig.~\ref{nmset3}.  Once again we find that the contribution of
$K^{*}$ exchange plays only a minor role.  We observe in average that
$\sigma_{np\rightarrow\Lambda\Theta^{+}_{-}}\sim$  5.0 $\mu b$ and   
$\sigma_{np\rightarrow\Sigma^{0}\Theta^{+}_{-}}\sim$  0.3 $\mu b$ in the
range of the CM energy $\sqrt{s}_{\rm th} \le \sqrt{s} \le  
3.5$ GeV. They are almost ten times smaller than those of
$\Theta^{+}_{+}$.  This behavior can be interpreted dynamically
by the fact that a large momentum transfer $\sim 800\, {\rm MeV}$
enhances the P-wave coupling of the $\Theta^{+}_{+}$ than the S-wave one
of the $\Theta^{+}_{-}$.

\begin{figure}[tbh]
\begin{tabular}{cc}
\resizebox{8cm}{5cm}{\includegraphics{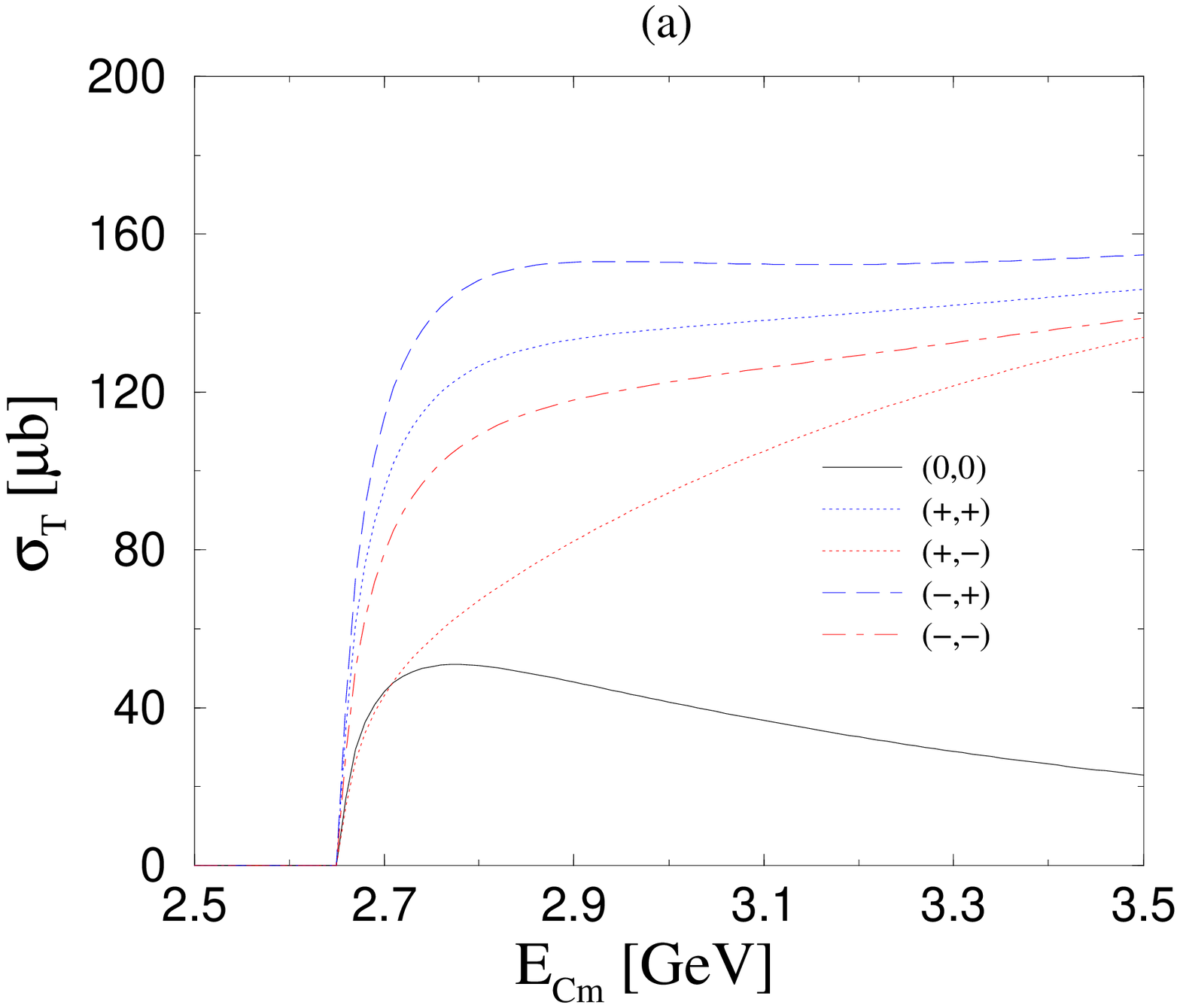}}
\resizebox{8cm}{5cm}{\includegraphics{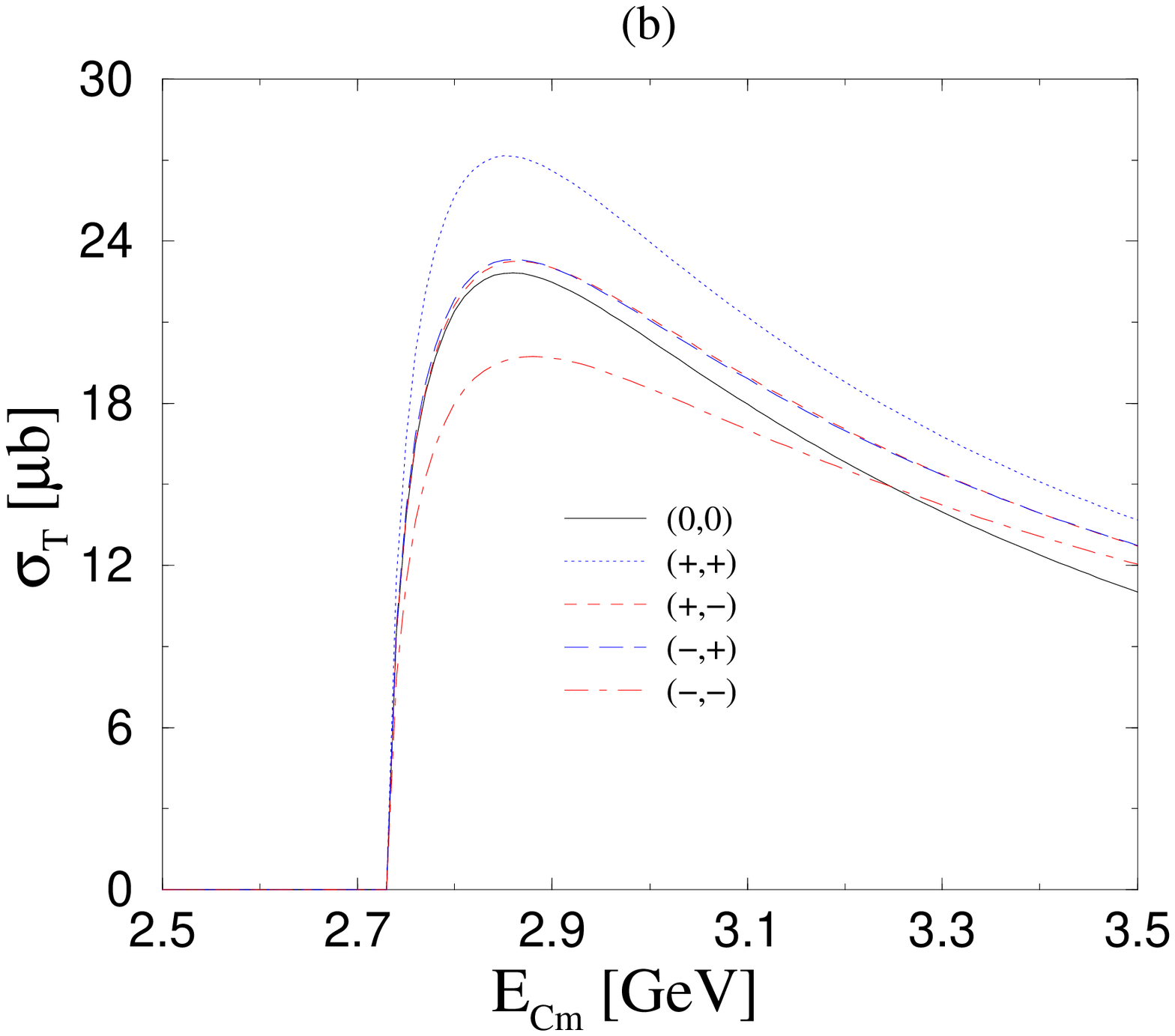}}
\end{tabular}
\caption{The total cross sections of
$np\rightarrow\Lambda\Theta^{+}_{+}$ in the left panel (a) and
$np\rightarrow\Sigma^{0}\Theta^{+}_{+}$ in the right panel (b).  The
parameter set of the J\"ulich--Bonn potential is employed.  The
notations are the same as in Fig.~\ref{nmset2}.}
\label{nmset4}
\end{figure}

In Fig.~\ref{nmset4}, we show the total cross sections of the
reactions for the $\Theta^{+}_{\pm}$ with the parameter set of the
J\"ulich--Bonn potential.  Here, different cutoff parameters are
employed at different vertices as mentioned previously.   
We find that the contribution from $K^*$ exchange turns out to be
larger in the $np\rightarrow\Lambda\Theta^{+}_{+}$ reaction
than in the $np\rightarrow\Sigma^{0}\Theta^{+}_{+}$.  This can be
easily understood from the fact that the J\"ulich--Bonn cutoff parameter
$\Lambda_{K^*N\Lambda}$ is chosen to be approximately twice as large as
that of the $KN\Lambda$ vertex, while the value of the
$\Lambda_{K^*N\Sigma}$ is about two times smaller than that of the
$\Lambda_{KN\Sigma}$.  The average total cross sections are obtained
as follows: $\sigma_{np\rightarrow\Lambda\Theta^{+}_{+}}\sim$  100
$\mu b$ and $\sigma_{np\rightarrow\Sigma^{0}\Theta^{+}_{+}}\sim$  20
$\mu b$ in the range of the CM energy $\sqrt{s}_{\rm th} \le \sqrt{s}
\le 3.5$ GeV.
  
\begin{figure}[tbh]
\begin{tabular}{cc}
\resizebox{8cm}{5cm}{\includegraphics{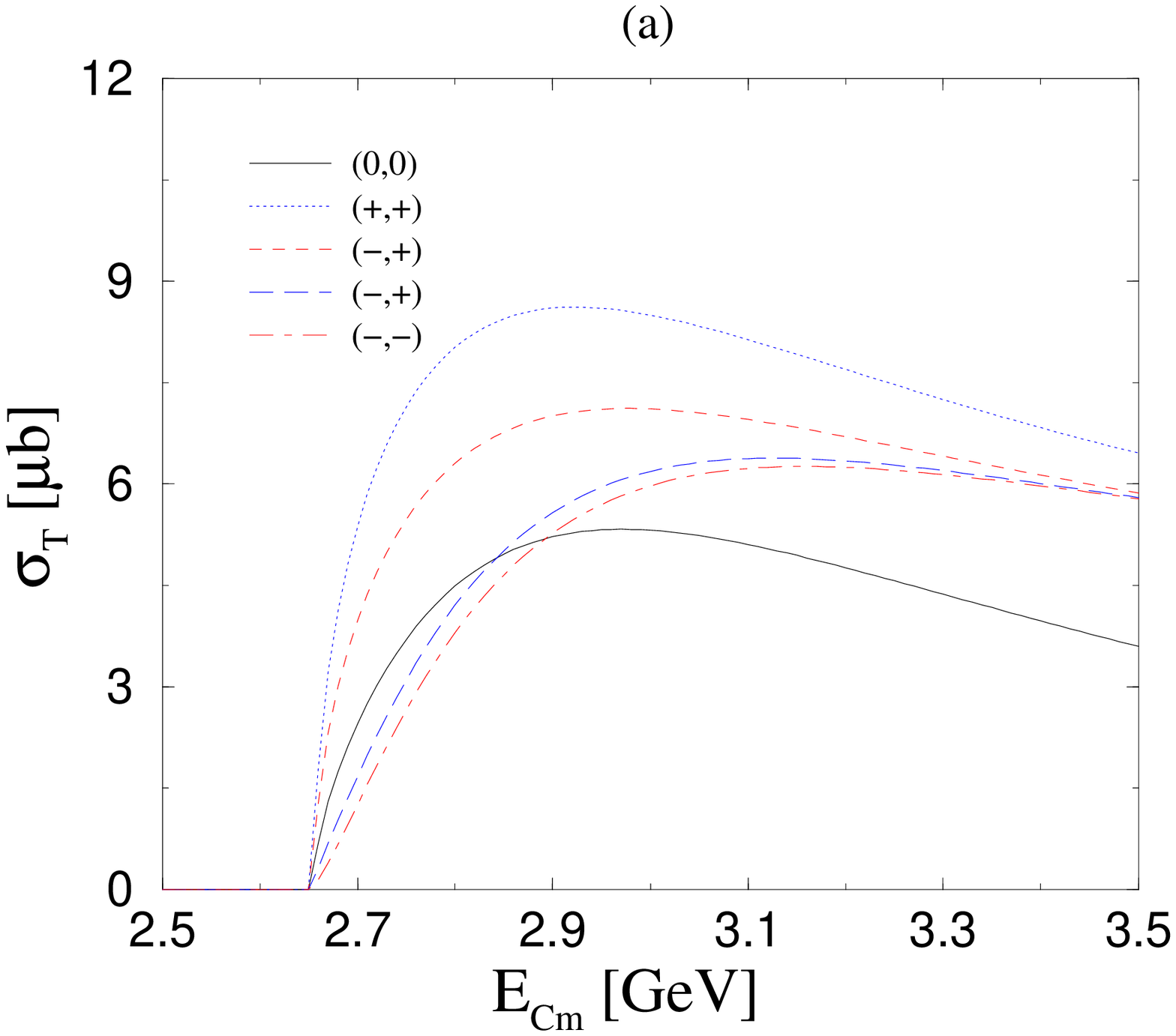}}
\resizebox{8cm}{5cm}{\includegraphics{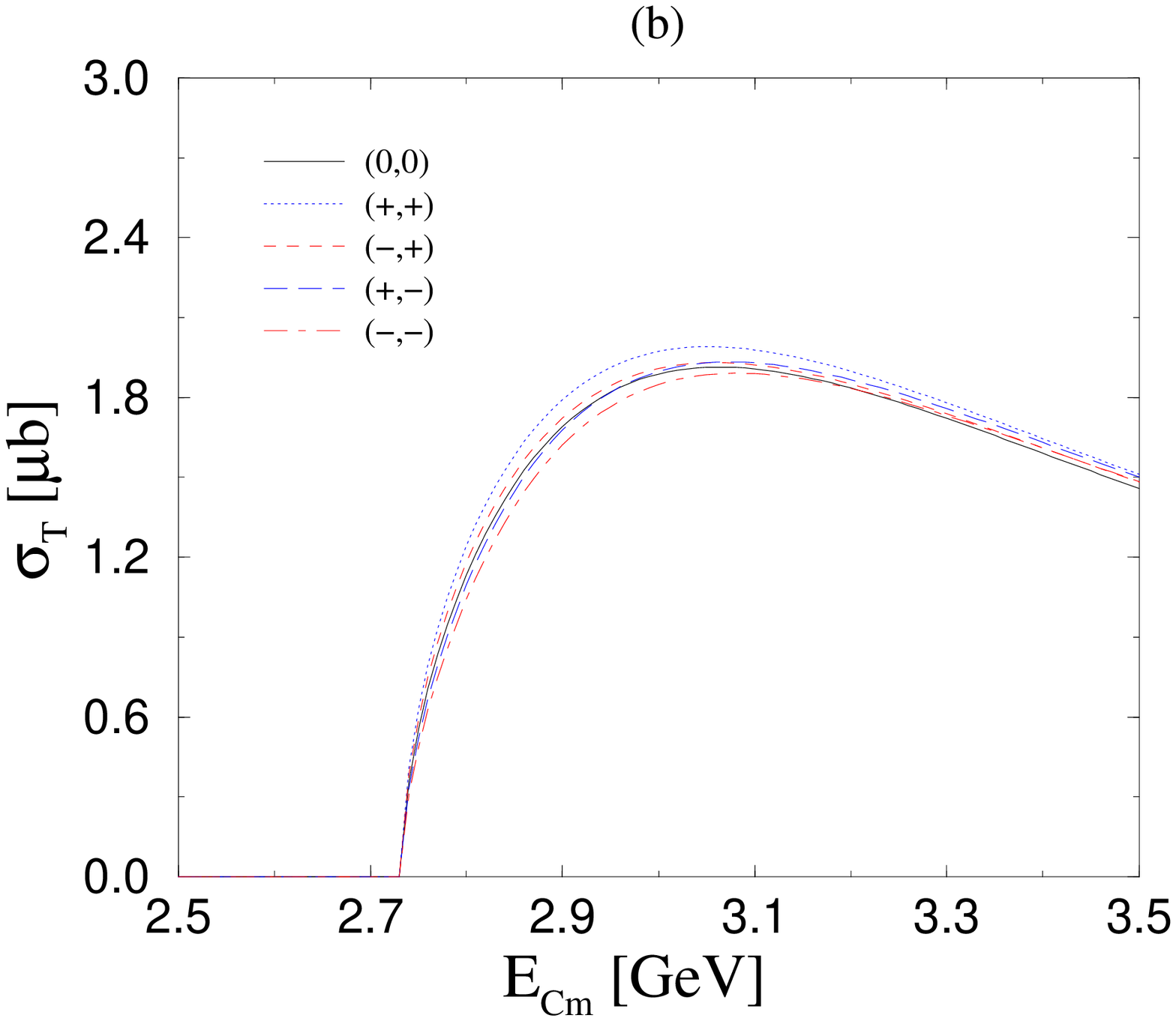}}
\end{tabular}
\caption{The total cross sections of
$np\rightarrow\Lambda\Theta^{+}_{-}$ in the left panel (a) and
$np\rightarrow\Sigma^{0}\Theta^{+}_{-}$ in the right panel (b).  The
parameter set of the J\"ulich--Bonn potential is employed.  The
notations are the same as in Fig.~\ref{nmset2}.}
\label{nmset5}
\end{figure}   
In Fig.~\ref{nmset5}, the total cross sections for $\Theta^{+}_{-}$ are
drawn.  In this case, the average total cross sections are given as
follows: $\sigma_{np\rightarrow\Lambda\Theta^{+}_{+}}\sim$ 
6.0 $\mu b$ and $\sigma_{np\rightarrow\Sigma^{0}\Theta^{+}_{+}}\sim$  2.0
$\mu b$ in the 
same range of the CM energy.  The results for the negative-parity
$\Theta_{-}^{+}$ are about fifteen times smaller than those of
$\Theta^{+}_{+}$.  

Compared to the results with the parameter set of the Nijmegen
potential, those with the J\"ulich--Bonn one are rather sensitive to
the signs of the coupling constants.  It is due to the fact that
the cutoff parameters taken from the J\"ulich--Bonn potential are
different at each vertex.  If we had taken similar values of the
cutoff parameters for the Nijmegen potential, we would have obtained
comparable results to the case of the J\"ulich-Bonn potential.

\section{Summary and Conclusion}
Motivated by a series of recent
works~\cite{Nakayama:2003by,Zhao:2003bm,
Yu:2003eq,Carlson:2003xb,Thomas:2003ak,Hanhart:2003xp,Nam:2004qy,
Mehen:2004dy,Rekalo:2004kb}, we
have studied the reactions $np\rightarrow\Lambda\Theta^{+}$ and 
$np\rightarrow\Sigma^{0}\Theta^{+}$, employing both of the negative and
positive parities for the $\Theta^+$.  We have considered
$K$ and $K^{*}$ meson exchanges in the Born approximation.  The
coupling constant for the $KN\Theta$ vertex has been fixed by using
experimental information on the width $\Gamma_{\Theta\rightarrow KN}$
as well as the mass $M_\Theta$, while those for $K^*$ exchange have
been estimated by using SU(3) symmetry~\cite{Liu:2003hu}. It turned
out that the contribution of $K$ exchange was dominant and that the
overall results were rather insensitive to the value of the cutoff
parameter $\Lambda$.  In conclusion, we have found that
$\sigma_{np\rightarrow   Y^{0}\Theta^{+}_{+}} >>  \sigma_{np\rightarrow 
  Y^{0}\Theta^{+}_{-}}$, as shown in Table.~\ref{table2} where we have
summarized 
the average total cross sections.   
\begin{table}[tbh]
\begin{tabular}{c|cccc}
 & \multicolumn{2}{c}{Nijmegen} & \multicolumn{2}{c}{J\"ulich-Bonn}
\\Final Hyperon & \hspace{1cm}$\Lambda$ \hspace{1cm} &
\hspace{1cm}$\Sigma^{0}$ \hspace{1cm}  
& \hspace{1cm}$\Lambda$ \hspace{1cm} & \hspace{1cm}$\Sigma^{0}$
\hspace{1cm} \\ \hline 
$\sigma_{P = +1}\,[\mu b]$ &
40 & 2.0 & 100 & 20\\ $\sigma_{P = -1}\,[\mu b]$ & 5.0 &
0.3 & 6.0 & 2.0
\end{tabular}
\caption{The average total cross sections}
\label{table2}
\end{table}

As suggested by Refs.~\cite{Zhao:2003bm,
Yu:2003eq,Thomas:2003ak,Hanhart:2003xp,Nam:2004qy,
Mehen:2004dy,Rekalo:2004kb}, the $NN$ interaction will provide a good
framework to determine the parity of the $\Theta^+$, though it might
still require an experimental challenge.  However, we anticipate that
we would provide a guideline together with recent works for future
experiments to pin down the parity of the $\Theta^+$.   
  
\section*{Acknowledgments}

HChK is grateful to J.K. Ahn, C.H. Lee, and I.K. Yoo for valuable
discussions and comments.  The work of HChK is supported by the
KOSEF grant R01\--2001\--00014 (2003).  He acknowledges in part the
support from the 21st COE Program  
``Towards A New Basic Science: Depth and Synthesis'' (Osaka
university).  The work of S.I. Nam has been
supported by the scholarship endowed from the Ministry of Education,
Science, Sports and Culture of Japan.  


\end{document}